\documentclass[preprint,groupedaddress,floatfix,%
nofootinbib]{revtex4}

\usepackage{euscript}
\usepackage{dcolumn}
\usepackage{graphicx}
\usepackage{epsfig}
\usepackage{hyperref}
\usepackage{graphicx}
\usepackage{dcolumn}
\usepackage{longtable}
\usepackage{times}
\usepackage{amsmath,amssymb,bm}
\usepackage[english]{babel}
\usepackage[latin1]{inputenc}
\usepackage{times}
\usepackage[T1]{fontenc}
\usepackage{color}
\usepackage{fancybox}
\usepackage{sidecap}
\usepackage{graphicx}
\usepackage{epsfig}
\usepackage{psfrag}
\usepackage{bbm}
\usepackage[english]{babel}
\usepackage[latin1]{inputenc}
\usepackage{times}
\usepackage[T1]{fontenc}
\usepackage{color}
\usepackage{fancybox}
\usepackage{sidecap}
\usepackage{psfrag}
\usepackage{bbm}
\usepackage{wasysym}
\usepackage{slashed}
\usepackage{tikz}
\usepackage{euscript}

\begin{document}

\title{Stabilizer Approximation II: From H$_2$O To C$_6$H$_6$}
\author{Jianan Wang}
\author{Chuixiong Wu}
\author{Fen Zuo\footnote{Email: \textsf{zuofen@miqroera.com}}}
\affiliation{Shanghai MiQro Era Digital Technology Ltd, Shanghai, China}

\begin{abstract}
We apply the stabilizer method to the study of some complicated molecules, such as water and benzene. In the minimal STO-3G basis, the former requires 14 qubits, and the latter 72 qubits, which is very challenging. Quite remarkably, We are still able to find the best stabilizer states at all the bond lengths. Just as the previously studied H$_2$, LiH and BeH$_2$ molecules, here the stabilizer states also approximate the true ground states very well, especially when the molecules are strongly distorted. These results suggest stabilizer states could serve as natural reference states when the system involves strong static correlation. And in the language of quantum computing, one would expect stabilizer states to be natural initial states for chemical simulation.
\end{abstract}
 \maketitle

\tableofcontents

\section{INTRODUCTION}

Modern computational chemistry crucially depends on the Hartree-Fock~(HF) approach~\cite{Hartree, Slater,Fock}. When the HF wavefunction provides a good approximation to the system, one could develop various perturbative methods to improve the accuracy to a rather high level. However, when the HF description deviates much from the true state, an accurate description of the system is usually difficult to achieve. For example, when a large molecule is distorted from the equilibrium configuration, many orbitals could become degenerate. The occupation of electrons in these degenerate orbitals will no longer be well described by a single HF state, but many HF states at the same time. To figure out these reference states in a huge Hilbert space will be a challenging task. Such a problem is recently illustrated in~\cite{TE22}.

Something similar happens in the recent application of quantum computing to chemical simulations. When applying variational algorithms to study the ground state of molecules, one usually takes the HF states, or some other product states, as the initial states. Around the equilibrium point this leads to quite nice results, just as in the classical treatment. But when the molecules are distorted strongly, the results start to deviate~\cite{HW22}. The reason is also similar to the classical case, though in a different language: the molecules could develop huge quantum entanglement in the dissociation procedure, which is not guaranteed in the shallow circuits commonly used.

In order to solve this problem, the key step is to prepare a nice reference/inital state with suitable entanglement. Such a state could be a nearly equal-weighted supposition of many product states, thus may be well approximated by stabilizers states~\cite{Gottesman96,Gottesman97}. According to the Gottesman-Knill theorem, such states can be efficiently prepared with Clifford circuits~\cite{Gottesman98}, if they are known in advance. However, if the required states are unknown, directly searching them using Clifford circuits could be very inefficient~\cite{Shang2021,CAFQA,CAFQA2}. An efficient way to find out the required stabilizer states, though heuristic, is to determine the stabilizers directly from the Pauli terms in the Hamiltonian. This is used long ago in the development of the toric code, in which the stabilizers encode the Hamiltonian and  the exact ground state by construction~\cite{Kitaev97}.

Recently we applied this approach to the chemical Hamiltonians, and found the required stabilizer states very quickly~\cite{SA-I} for the small molecules such as H$_2$, LiH and BeH$_2$. In this paper we try to extend this approach to more complicated molecules, to see if the stabilizer states could still be efficiently obtained. According to the study in~\cite{CAFQA}, the Clifford-circuit search is already difficult for the water molecule, and some ambiguous results have been reported. As for benzene, a complete description is beyond the ability of current quantum hardware and simulators, so the frozen-core approximation is usually taken~\cite{TE22}. Therefore, we use these two cases to test the capability and efficiency of the stabilizer method.

\section{THE CALCULATION}

\subsection{H$_2$}

First we repeat the analyses for the hydrogen molecule~\cite{SA-I} to illustrate the procedure, but choose the Jordan-Wigner~(JW) transformation instead of the parity one~\cite{Parity}. At equilibrium point $d=0.74$\AA, the qubit-form Hamiltonian reads:
\begin{eqnarray}
H_0&=& -0.812 * \bar I\bar III\nonumber\\
&&- 0.223 * \bar Z\bar III\nonumber\\
&&- 0.223 * \bar I\bar IZI\nonumber\\
&&+ 0.174 * \bar Z\bar IZI\nonumber\\
&&+ 0.171 * \bar I\bar ZII\nonumber\\
&&+ 0.171 * \bar I\bar IIZ\nonumber\\
&&+ 0.169 * \bar I\bar ZIZ\nonumber\\
&&+ 0.166 * \bar Z\bar IIZ\nonumber\\
&&+ 0.166 * \bar I\bar ZZI\nonumber\\
&&+ 0.121 * \bar I\bar IZZ\nonumber\\
&&+ 0.121 * \bar Z\bar ZII\nonumber\\
&&+ 0.045 * \bar X\bar XXX\nonumber\\
&&+ 0.045 * \bar Y\bar YYY\nonumber\\
&&+ 0.045 * \bar X\bar XYY\nonumber\\
&&+ 0.045 * \bar Y\bar YXX.
\end{eqnarray}
We use $X,Y,Z,I$ for the gates acting on the spin-down orbitals, and $\bar X,\bar Y,\bar Z, \bar I$ for the gates on the spin-up ones. To approximate the ground state, one must choose the stabilizers in such a way that the corresponding energy is as low as possible. The numerical coefficients of the Pauli terms above suggest the following choice:
\begin{equation}
\bar Z\bar III, \bar I\bar IZI, -\bar I\bar ZII, -\bar I\bar IIZ.
\end{equation}
The corresponding state is the Hatree-Fock state $|01;01\rangle$, and the corresponding energy can be obtained by inserting the stabilizing condition back into the Hamiltonian.

When the molecule is stretched, the coefficients in the Hamiltonian change. For example, at $d=2.8$\AA, the Hamiltonian becomes:
\begin{eqnarray}
H_1&=&-0.734 * \bar I\bar III\nonumber\\
&&+ 0.122 * \bar Z\bar IZI\nonumber\\
&&+ 0.120 * \bar Z\bar IIZ\nonumber\\
&&+ 0.120 * \bar I\bar ZZI\nonumber\\
&&+ 0.119 * \bar I\bar ZIZ\nonumber\\
&&+ 0.073 * \bar X\bar XYY\nonumber\\
&&+ 0.073 * \bar Y\bar YXX\nonumber\\
&&+ 0.073 * \bar X\bar XXX\nonumber\\
&&+ 0.073 * \bar Y\bar YYY\nonumber\\
&&+ 0.048 * \bar I\bar IIZ\nonumber\\
&&+ 0.048 * \bar I\bar ZII\nonumber\\
&&+ 0.047 * \bar I\bar IZZ\nonumber\\
&&+ 0.047 * \bar Z\bar ZII\nonumber\\
&&+ 0.031 * \bar I\bar IZI\nonumber\\
&&+ 0.031 * \bar Z\bar III.
\end{eqnarray}
An intuitive choice for the stabilizers is
\begin{equation}
-\bar Z\bar IZI, \quad -\bar Z\bar IIZ,\quad -\bar I\bar ZZI.
\end{equation}
When these are selected, the Hamiltonian could be completely simplified. So the stabilizer states are doubly degenerate. To fix the states explicitly, one may choose one more stabilizer as either $\bar X\bar XYY$ or $-\bar X\bar XYY$. Then the correspond states read
\begin{equation}
\frac{|11;00\rangle\pm|00;11\rangle}{\sqrt{2}}. \label{eq.gGHZ}
\end{equation}
These are variants of the 4-qubit Greenberger-Horne-Zeilinger~(GHZ) state. We will write $n$ repeating bits as $1^n$ or $0^n$  for short. Actually the degeneracy destroys the entanglement, and one could represents them as product states. This can be done by choosing the last stabilizer as $\pm \bar I\bar IIZ$, and get:
\begin{equation}
|1^2;0^2\rangle, \quad |0^2;1^2\rangle.\label{eq.hs}
\end{equation}
Now these states are of spin 1, and different resonating forms of them give (\ref{eq.gGHZ}). The appearance of such high-spin states should be expected, according to the Hund's rule~\cite{Hund}.

A careful analysis of the above Hamiltonian reveals that there is another degenerate state, with the stabilizers:
\begin{equation}
-\bar Z\bar IZI,\quad -\bar I\bar ZIZ, \quad -\bar X\bar XYY, \quad -\bar I\bar IZZ.
\end{equation}
And the state reads:
\begin{equation}
\frac{|10;01\rangle-|01;10\rangle}{\sqrt{2}}.\label{eq.ls}
\end{equation}
The reason why such a state could achieve an energy as low as (\ref{eq.hs}) could be explained by the interference effect of the superposition, or resonating~\cite{Pauling}. As the resonating is between two different spin sectors, we call it a ``spin-resonating state". This also explains why the other combination, with a plus sign inbetween, has a higher energy.
Notice that for all the states (\ref{eq.gGHZ},\ref{eq.hs},\ref{eq.ls}), the two spatial orbitals are half-filled, in agreement with the asymptotic behavior of full configuration interaction (FCI) results~\cite{HJO}.

\begin{figure}[ht]
\centering
	\includegraphics[width=0.8\textwidth]{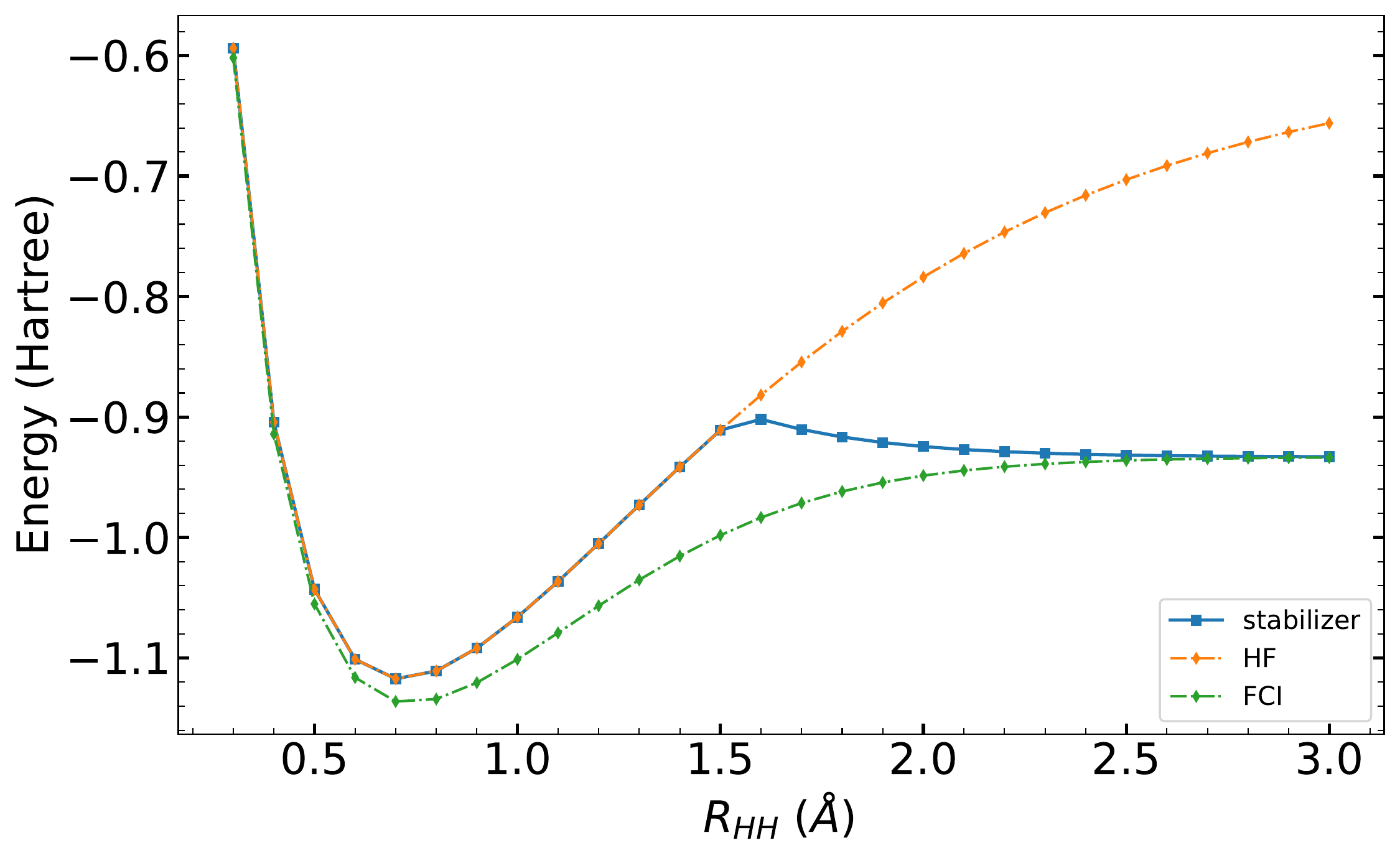}
\caption{\it Dissociation Curve of H$_2$. The results from the stabilizer method are plotted together with those from HF and FCI.}\label{fig:H2}
\end{figure}

Repeating this procedure for all the bond lengths, we obtain the approximate potential energy curve. The results are plotted in Fig.~\ref{fig:H2}, together with those from HF and FCI. As emphasized in~\cite{CAFQA} and \cite{SA-I}, the stabilizer method behaves correctly in the dissociation limit, and thus recover almost the whole correlation energy missed in the HF approximation.

\newpage
\subsection{H$_2$O}
For H$_2$O there are 7 atomic orbitals involved, and we need 14 qubits to describe them. After JW transformation the Hamiltonian contains about 3000 Pauli terms. More than half of these terms have coefficients smaller than $10^{-6}$ Hartree, and are then neglected in previous studies. We focus on the dissociation with the H-O-H bond angle fixed at equilibrium value $105^{\circ}$.

Around the equilibrium point, the stabilizer method reproduces the HF state, in which 5 spatial orbitals are fully occupied, and the other 2 are empty. Explicitly, the state is
\begin{equation}
|0^21^5;0^21^5\rangle.
\end{equation}

When the O-H bond is stretched, the lowest 3 spatial orbitals are unaffected. For simplicity, We will omit them in the following expressions. The higher 4 orbitals start to change when the O-H distance goes beyond about 1.5\AA. Same as the hydrogen molecule case, they become half-filled eventually. This is also in agreement with the FCI results~\cite{HJO}. Again we have different spin states. First we will immediately get the high spin states:
\begin{equation}
|1^4;0^4\rangle,\quad |0^4;1^4\rangle. \label{eq.S2}
\end{equation}
Further analysis of the Hamiltonian shows that there are also spin-resonating states, determined by the stabilizers:
\begin{eqnarray}
&&-\bar I\bar I\bar I\bar ZIIIZ,\quad -\bar I\bar I\bar Z\bar IIIZI, \quad -\bar I\bar Z\bar I\bar IIZII, \quad -\bar Z\bar I\bar I\bar IZIII,\nonumber\\
&&-\bar I\bar I\bar X\bar XIIYY,\quad -\bar I\bar X\bar X\bar IIYYI, \quad -\bar X\bar X\bar I\bar IYYII.
\end{eqnarray}
Now we have 7 stabilizers for 8 qubits, which means a doubly degenerate state. The degeneracy could be lifted by adding an extra stabilizer, which we choose to be $\pm \bar I\bar I\bar I\bar IZZZZ$. With $-\bar I\bar I\bar I\bar IZZZZ$ included, we obtain an odd-spin state:
\begin{eqnarray}
&&\left(+|1000;0111\rangle+|1011;0100\rangle-|0100;1011\rangle-|0111;1000\rangle\right.\nonumber\\
&&\left.~+|0010;1101\rangle-|0001;1110\rangle-|1101;0010\rangle+|1110;0001\rangle\right)/2\sqrt{2}.\label{eq.S-odd}
\end{eqnarray}
If choosing $\bar I\bar I\bar I\bar IZZZZ$ instead, we get the following even-spin state:
\begin{eqnarray}
&&\left(+|1010;0101\rangle-|0110;1001\rangle-|1100;0011\rangle-|0000;1111\rangle\right.\nonumber\\
&&\left.~-|1001;0110\rangle+|0101;1010\rangle-|0011;1100\rangle-|1111;0000\rangle\right)/2\sqrt{2}.\label{eq.S-even}
\end{eqnarray}

One should notice that these states do not have conserved numbers of spin-up and spin-down electrons. However, since (\ref{eq.S-even}) is degenerate with the high-spin ones (\ref{eq.S2}), we could make a new combination of them to recover the conservation of spin-up and spin-down electrons. The resulting state will have total spin 0. In practice, these low-spin stabilizer states (\ref{eq.S-odd},\ref{eq.S-even}) are not easy to identify, in contrast to the high-spin ones~(\ref{eq.S2}). This is similar to the situation in the Clifford-circuit approach, in which the lowest singlet state for H$_2$O is never found actually~\cite{CAFQA}.

\begin{figure}[ht]
\centering
	\includegraphics[width=0.8\textwidth]{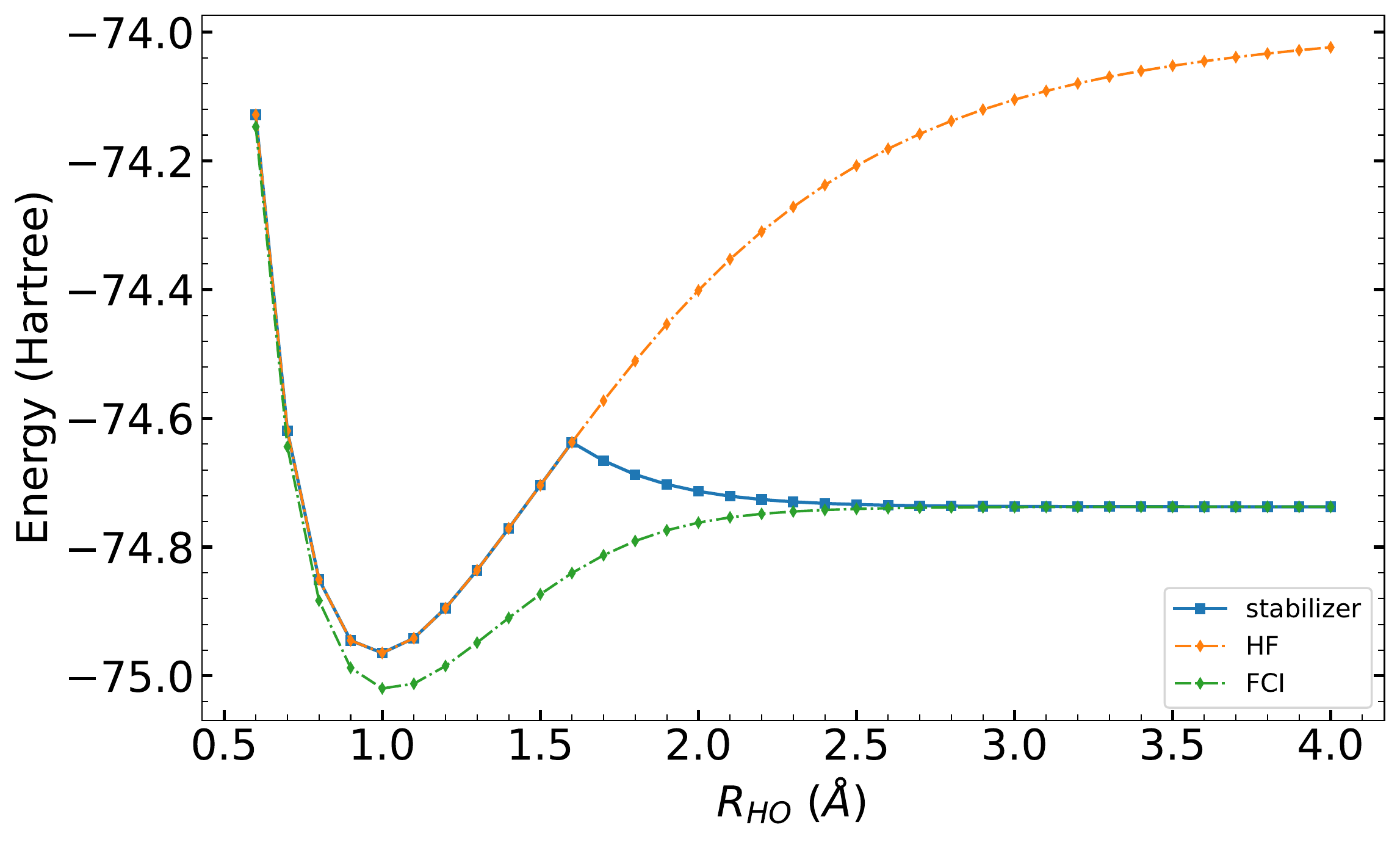}
\caption{\it Dissociation Curve of H$_2$O at fixed bond angle. The results from the stabilizer method are compared with those from HF and FCI.}\label{fig:H2O}
\end{figure}

The corresponding dissociation curve is plotted in Fig.~\ref{fig:H2O}, together with the HF and FCI results. The correct dissociation limit is clearly recovered again. Moreover, the stabilizer method gives a single degenerate curve, in contrast to those from the Clifford-circuit search~\cite{CAFQA}. In \cite{HW22} it is pointed out that variational algorithms with initial product states fail to give accurate results for stretched molecules, including H$_2$O. Fig.~\ref{fig:H2O} shows that stabilizer states may help improving the accuracy in such cases, if taken as initial states.

\newpage
\subsection{C$_6$H$_6$}
Now we turn to a challenging system, the benzene molecule. It involves 42 electrons and 36 spatial orbitals, so we need 72 qubits to describe them. And the Hamiltonian contains about 2.5 million Pauli terms. This makes it very difficult to treat the whole system directly. Previous study utilizes the frozen-core approximation, and chooses active spaces with 8-16 qubits~\cite{TE22}. This is in some sense comparable to the traditional H\"{u}ckel approximation~\cite{Huckel}. Here we try to go beyond such an approximation, utilizing the stabilizer method.

\subsubsection{H\"{u}ckel case}

First we consider the situation when only the C-C bonds are stretched, keeping the hexagonal symmetry. This is recently studied in detail in~\cite{TE22}. Due its similarity with the H\"{u}ckel picture~\cite{Huckel}, we call it H\"{u}ckel case.

At equilibrium, we expect 21 of the total 36 spatial orbitals would be fully occupied, and get the HF state:
\begin{equation}
|0^{15}1^{21};0^{15}1^{21}\rangle.
\end{equation}
When the C-C bonds are stretched, the 6 active electrons in the delocalized $\pi$ orbitals will first be excited. The other 36 electrons may stay paired up in the remaining orbitals. Thus we should get a state with 18 fully-occupied orbitals, 6 half-filled ones, and 12 empty ones. This is indeed the case. For the half-filled orbitals, the 6 active electrons could take the same spin orientation and give rise to the high-spin state:
\begin{equation}
|1^6;0^6\rangle, \quad |0^6;1^6\rangle.~\label{eq.benzene-6}
\end{equation}
Alternatively, they could hop between orbitals with different spin orientation, leading to the spin-resonating state. This is described by the following stabilizers:
\begin{eqnarray}
&&-\bar Z\bar I\bar I\bar I\bar I\bar IZIIIII,\nonumber\\
&&-\bar I\bar Z\bar I\bar I\bar I\bar IIZIIII,\nonumber\\
&&-\bar I\bar I\bar Z\bar I\bar I\bar IIIZIII,\nonumber\\
&&-\bar I\bar I\bar I\bar Z\bar I\bar IIIIZII,\nonumber\\
&&-\bar I\bar I\bar I\bar I\bar Z\bar IIIIIZI,\nonumber\\
&&-\bar I\bar I\bar I\bar I\bar I\bar ZIIIIIZ,\nonumber\\
&&-\bar X\bar X\bar I\bar I\bar I\bar IYYIIII,\nonumber\\
&&-\bar I\bar X\bar X\bar I\bar I\bar IIYYIII,\nonumber\\
&&-\bar I\bar I\bar X\bar X\bar I\bar IIIYYII,\nonumber\\
&&-\bar I\bar I\bar I\bar X\bar X\bar IIIIYYI,\nonumber\\
&&-\bar I\bar I\bar I\bar I\bar X\bar XIIIIYY.\label{eq.benzene-33}
\end{eqnarray}
The corresponding state is a little complicated, and will not be shown explicitly here. It is doubly degenerate, and also degenerate with the high-spin states (\ref{eq.benzene-6}).

\begin{figure}[ht]
\centering
	\includegraphics[width=\textwidth]{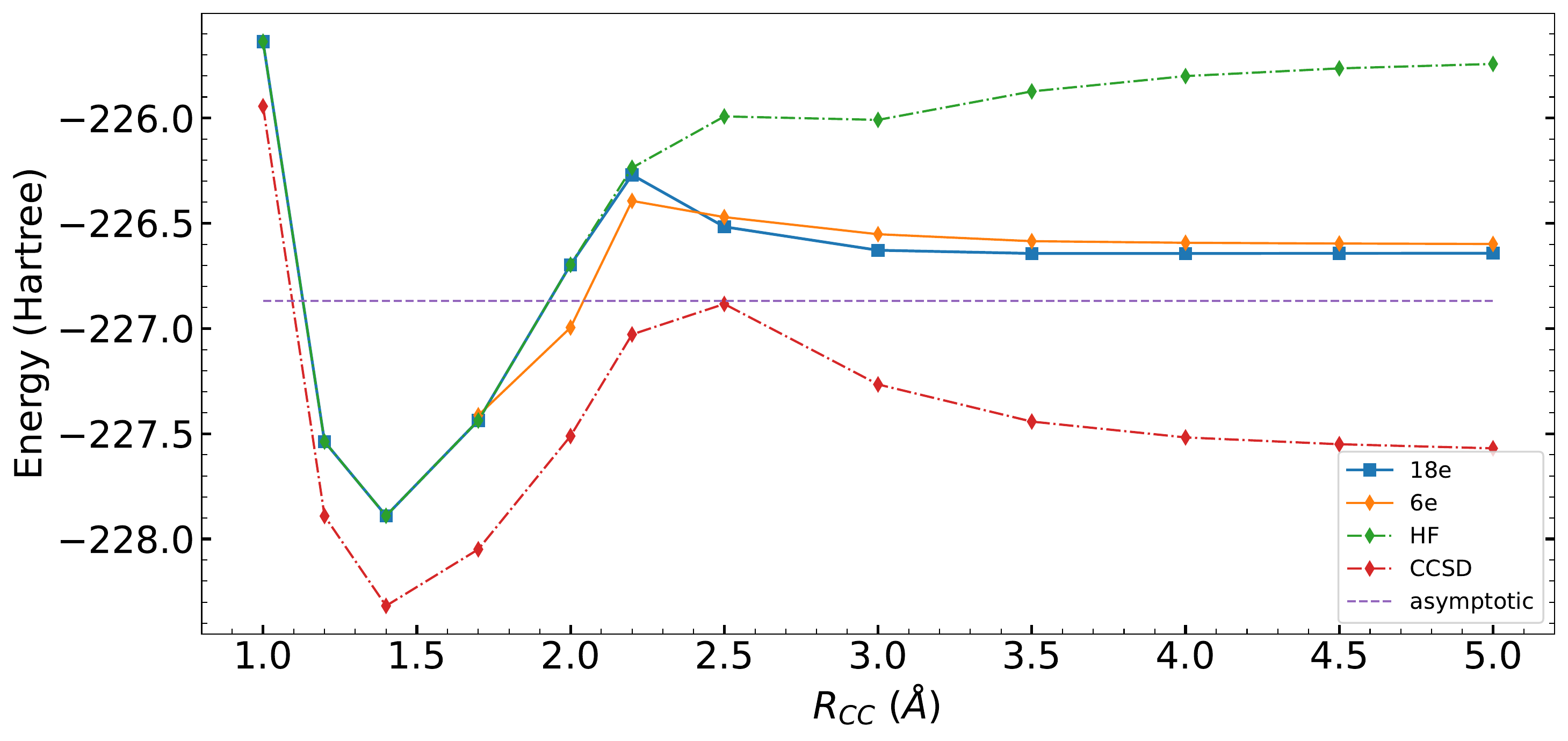}
\caption{\it Potential energy curve of benzene dissociated symmetrically into 6 free CH radicals. The two curves labeled by ``6e" and ``18e" correspond to different stabilizer states, and are explained in the contexts. The blue dash line labeled by ``asymptotic" denotes the total energy of the radicals, calculated with FCI and the STO-3G basis. }\label{fig:Huckel}
\end{figure}

The appearance of 6 unpaired electrons has been numerically indicated in recent study~\cite{TE22}. In our calculation, we find another stabilizer state with even lower energy, at least in the asymptotic region. This state has 12 fully-occupied orbitals, 18 half-filled ones, and 6 empty ones. We will investigate the details of this state later. Therefore, the frozen-core approximation fails to give the right physical picture here. This is not surprising: the resulting CH radicals are not stable, and contain active electrons. The comparison of the energy profiles calculated from different methods are shown in Fig.~\ref{fig:Huckel}. Since FCI is not available anymore, we provide the results calculated with the Coupled-Cluster Singles-and-Doubles~(CCSD) method. As a whole, our results from the stabilizer approximation resemble and improve the ``full diagonalization of the active-space Hamiltonian" results in~\cite{TE22}.

We also notice that the stabilizer curves do not approach the dissociation limit accurately. This is reasonable, since the HF/stabilizer description of the CH radical at fixed bond length is not accurate enough.

\subsubsection{Overall case}

To probe the state with 18 active electrons, we dissociate the benzene completely into individual atoms. This is done by stretching the C-C bonds and C-H bonds simultaneously, keeping their ratios fixed.

Again, the 18 electrons could occupy spin orbitals of the same orientation and lead to the high-spin states:
\begin{equation}
|1^{18};0^{18}\rangle,  \quad    |0^{18};1^{18}\rangle \label{eq.benzene18}
\end{equation}
Among them, the 6 electrons from the delocalized $\pi$ orbital could further hop between orbitals with different spin orientations, and gives the spin-resonating state (\ref{eq.benzene-33}). As for the remaining 12 unpaired electrons, it would be convenient to consider them from the atomic orbital point of view. Then one could guess that 6 electrons would occupy the $1s$ orbitals of the H atoms, and the other 6 occupy the $2p$ orbitals of the C atoms. Since there are 6 $2p_x$ orbitals and 6 $2p_y$ orbitals, the 6 electrons may hop between them. Indeed, for these 6 electrons we get the spatial-resonating state:
\begin{eqnarray}
&&-ZIIIIIZIIIII,\nonumber\\
&&-IZIIIIIZIIII,\nonumber\\
&&-IIZIIIIIZIII,\nonumber\\
&&-IIIZIIIIIZII,\nonumber\\
&&-IIIIZIIIIIZI,\nonumber\\
&&-IIIIIZIIIIIZ,\nonumber\\
&&+XXIIIIXXIIII,\nonumber\\
&&+IXXIIIIXXIII,\nonumber\\
&&+IIXXIIIIXXII,\nonumber\\
&&+IIIXXIIIIXXI,\nonumber\\
&&+IIIIXXIIIIXX.\label{eq.benzene-06}
\end{eqnarray}
The explicit form of the stabilizers differ sightly from those of (\ref{eq.benzene-33}). All the 12 electrons still remain in the same spin sector, as required by Hund's rule. One may expect some fully-entangled state of the total 18 electrons, but we do not find such states yet.

\begin{figure}[ht]
\centering
	\includegraphics[width=\textwidth]{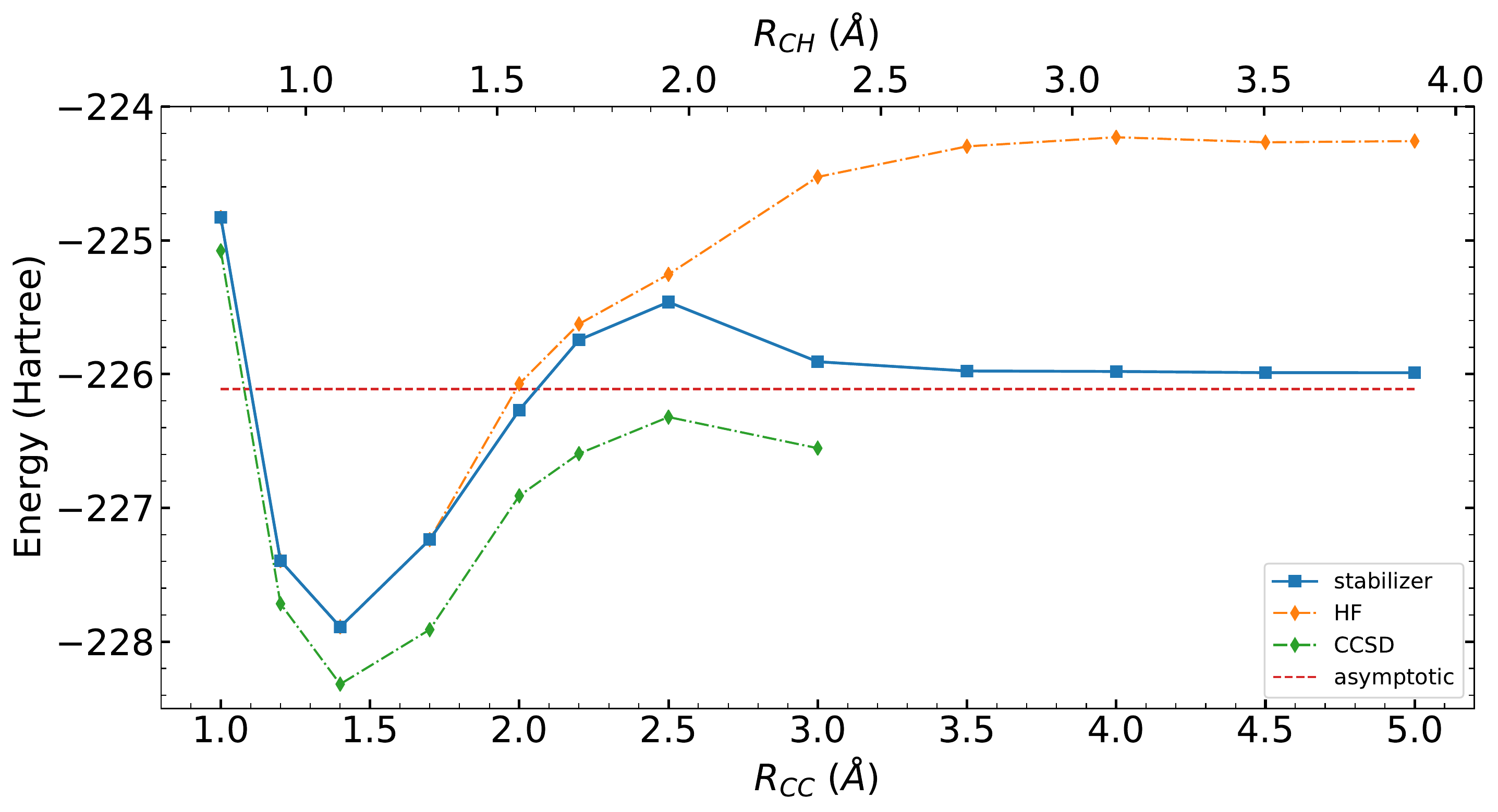}
\caption{\it Potential energy curve of benzene dissociated symmetrically into individual C and H atoms. The red dash line labeled by ``asymptotic" denotes the total energy of the atoms after dissociation, calculated with FCI and the STO-3G basis.}\label{fig:Overall}
\end{figure}

It turns out the high-spin states (\ref{eq.benzene18}), the one with spin resonating(\ref{eq.benzene-33}), and the one with both spin resonating(\ref{eq.benzene-33}) and spatial resonating (\ref{eq.benzene-06}) all have nearly the same energy in the asymptotic region. In Fig.~\ref{fig:Overall} we plot the dissociation curves calculated from different methods. For the results from the stabilizer method, all the mentioned states fall on a single curve. remarkably, the stabilizer method nearly reproduces the asymptotic result. The deviation from the asymptotic line is about $0.12$ Hartree when the C-C distance is $5.0$\AA. We are unable to obtain the CCSD results in the large distance region due to appearance of numerical singularities.

\newpage
\section{DISCUSSION}

We have demonstrated the efficiency and efficacy of the stabilizer method with two examples, the water and benzene molecules. Even for the challenging benzene molecule, we are able to find the proper stabilizers quickly for all the molecular configuration. In contrast to the HF method, the stabilizer states always stay close to the true ground states. Therefore, they could serve as proper reference/initial states for further refined calculations, either for classical methods or quantum simulation. In particular, it would be interesting to check if they qualify as suitable initial states for the quantum phase estimation algorithm, to guarantee the exponential quantum advantage in quantum chemistry~\cite{GKC2208}.

To confirm the results obtained in this paper, we need to apply the method to even more complicated systems, for example, to molecules containing transition metal atoms. Recently the chromium dimer has already been studied with Clifford-circuit search~\cite{CAFQA}. It would be interesting to investigate such system directly with stabilizers.

States from the JW transformation have a direct interpretation on the chemistry side. In studying the water molecule we find that the stabilizer states do not preserve electrons with individual spin orientations. Then what would these states present themselves with the parity and Bravyi-Kitaev transformation~\cite{BK}? Some preliminary study has been performed recently with Clifford circuits~\cite{CAFQA2}. The stabilizer analysis may provide alternative answers.

Many combinational optimization problems can be reduced to a groundstate-finding problem for a classical Hamiltonian. In~\cite{CAFQA2} it is concluded that Clifford circuits may not help solving such problems. However, from the stabilizer point of view, these classical Hamiltonians should be even easier to solve. Moreover, stabilizer states could provide the exact solution to such classical problems, not just an approximation.

We leave all these for future study.

\section*{Framework and Resource}

The Hartree-Fock, Coupled-Cluster Singles-and-Doubles, and Full Configuration Interaction calculation are performed using Python-based Simulations of Chemistry Framework (PySCF)~\cite{PySCF}. We develop original submodules to implement the Jordan-Wigner transformation and the stabilizer-finding process.

All the calculation has been done using a desktop computer with a 8-core CPU and 64 GB RAM. With the stabilizer method, a single-point calculation of the benzene molecule can be finished within 10 minutes.



\end{document}